# The Observation of Ferroelastic and Ferrielectric Domains in AgNbO$_3$ Single Crystal

Wei Zhao[1], Zhengqian Fu[2]*, Jianming Deng[1], Song Li[3], Yifeng Han[4], Man-Rong Li[4], Xueyun Wang[1]*, Jiawang Hong[1]*

[1]School of Aerospace Engineering, Beijing Institute of Technology, Beijing 100081, China

[2]The State Key Lab of High Performance Ceramics and Superfine Microstructure, Shanghai Institute of Ceramics, Chinese Academy of Sciences, Shanghai 200050, China

[3]Key Laboratory of Inorganic Functional Materials and Devices, Shanghai Institute of ceramics, Chinese Academy of Sciences, Shanghai 200050, China

[4]Key Laboratory of Bioinorganic and Synthetic Chemistry of Ministry of Education, School of Chemistry, Sun Yat-Sen University, Guangzhou 510275, China

*Corresponding author. Email: fmail600@mail.sic.ac.cn; xueyun@bit.edu.cn; hongjw@bit.edu.cn

**ABSTRACT**

Compared to AgNbO$_3$ based ceramics, the experimental investigations on the single crystalline AgNbO$_3$, especially the ground state and ferroic domain structures, are not on the same level. Here in this work, based on successfully synthesized AgNbO$_3$ single crystal using flux method, we observed the coexistence of ferroelastic and ferrielectric domain structures by a combination study of polarized light microscopy and piezoresponse force microscope, this finding may provide a new aspect for studying AgNbO$_3$. The result also suggests a weak electromechanical response from the ferrielectric phase of AgNbO$_3$ which is also supported by the transmission electron microscope characterization. Our results reveal that the AgNbO$_3$ single crystal is in a polar ferrielectric phase at room temperature, clarifying its ground state which is controversial from the AgNbO$_3$ ceramic materials.

Silver niobate, AgNbO$_3$ (ANO), with ABO$_3$-type perovskite structure has pulled significant scientific research interests owing to its lead-free and high energy storage capacity.[1] Up to now, the recoverable energy density reaches 6.3 J/cm$^3$,[2] due to its antiferroelectric nature, leading to the enhancement of recoverable energy density $W_{rec}$.[3] There are many different strategies to improve the energy storage performance of ferroic materials.[4–6] A comprehensive studies trying to modify or reshape the hysteresis loop for larger storage capacity have been demonstrated through A-site doping,[7–12] B-site doping[13,14] and simultaneously doping both sites.[15] However, there exists nontrivial and critical debates about the crystalline structure of ANO at room temperature.[1,16] Nevertheless, the crystalline structure is imperative for understanding the underlying physics for the manipulation of antiferroelectric phase, which may in turn boost the research for energy storage enhancement.

Since ANO has been discovered in 1958,[17] the crystalline structure at room temperature is in debate. Plenty of works suggest the existence of either *Pbcm* or *Pmc*2$_1$ space group from both sophisticated experiments (X-Ray Diffraction, Transmission Electron Microscope, Neutron Diffraction, etc.) and comprehensive theoretical calculations,[16–24] but the debate is not resolved, mainly due to the undistinguishable diffraction patterns of those two structures and a relatively small energy difference (0.1-0.5 meV/f.u) between these two structures.[25,26] Most works show that *Pbcm* describes the structure better at room temperature, which is relative to the antiferroelectric phase.[1,8,11,14] Note that a ferrielectric phase with *Pb*2$_1$*m* (***b c a*** settings of *Pmc*2$_1$, we use *Pmc*2$_1$ for later discussion)[24,27] structure is also proposed, in which Ag and Nb exhibit antiparallel displacements with a net polarization along the *c*-axis.[21,23] Note that *Pbcm* is centrosymmetric and *Pmc*2$_1$ is non-centrosymmetric, and non-centrosymmetric structure usually gives rise to ferroic (including ferroelastic and ferroelectric) domain. Ferroic domains are very important in both theoretical research and practical application.[28–30] Therefore, revealing the existence of possible ferroic domain may solve the debate on the crystalline structure of ANO.[31,32] Hitherto, there are limited reports on the domain structure of ANO, and also restricted to the transmission electron microscope technique.[18,20,21,33] In addition, the domain structure related works are based on polycrystalline specimens, in which the grain boundary, strain or strain gradient *etc*. give non-negligible influences on the crystal structures.[34–37] Therefore, synthesis the ANO single crystals and investigation of the domain structure are critical.

In this work, based on the synthesized single crystal ANO, we use polarized light microscopy

(PLM) and piezoresponse force microscope (PFM) to reveal that ferroelastic and ferrielectric domains coexist in single crystalline ANO. Transmission electron microscope (TEM) results also suggest ANO is in ferrielectric phase with space group $Pmc2_1$ at room temperature with the existence of antiphase domain boundaries.

The ANO single crystals were grown by a flux method. $Ag_2O$ (99.9%), $Nb_2O_5$ (99.99%), and $V_2O_5$ (99.2%) with molar ratio 7.4:1:4 was used as described elsewhere.[38,39] Note that $Ag_2O$ and $V_2O_5$ are both as flux. The mixture was well grinded and put into an alumina crucible, and then heated to 1255 K with 100 K/h heating rate, after a melting and dissolving process at 1255 K for 5 h, the mixture was cooled to 1050 K with a rate of 1 K/h, followed by a 100 K/h cooling to room temperature. $HNO_3$ and NaOH solution were used as etchant to remove the flux matrix.[39] Brownish crystals were obtained. ANO ceramic samples were also prepared using sintering method.[1,13] Raw materials were well grounded with stoichiometric mixture of $Nb_2O_5$ and $Ag_2O$, then the mixture was calcined at 900 °C for 10 h. The calcined powders were milled and pressed into pellets with a thickness of 1.5 mm and diameter of 10 mm, the pellets were sintered at 1000 °C for 10 h. The sample quality and crystal structure were identified on crashed powder of single crystalline ANO and ceramic powders by a Bruker D8 Advance X-ray diffractometer using Cu $K\alpha$ radiation. The TEM investigation was carried out on JEOL JEM-2100F microscope operated at 200 kV. TEM specimens were prepared by a conventional approach of mechanical thinning and finally $Ar^+$ milling in a Gatan PIPS II. The ion-milling voltage was gradually decreased from 3 keV to 0.5 keV to reduce ion-beam damage. Specimens were then coated with a thin layer of carbon to minimize charging under the electron beam. Cleaved surface with ferroic domain structure was observed by using polarized microscopy (Zeiss Optical Microscope Axio Image) with reflection-mode. PFM measurement was carried out by using Asylum Research MFP-3D at room temperature. The sample was grounded on conductive substrate. 0.5-1 V were applied to the Pt/Ir-coated Silicon tip during the PFM measurement.

As grown ANO single crystal sample is characterized by X-ray diffraction. Fig. 1 shows the X-ray diffraction results of grinded powders of ANO single crystals and polycrystalline ANO at room temperature. Based on space group of $Pbcm$ and $Pmc2_1$, all the observed peaks could be indexed. The diffraction patterns agree well with those reported for ANO (orthorhombic, $Pbcm$ or $Pmc2_1$).[1,19,21] There is no obvious $Ag_2O$ or $V_2O_5$ flux left over, due to the complete removal of flux from chemical etching.

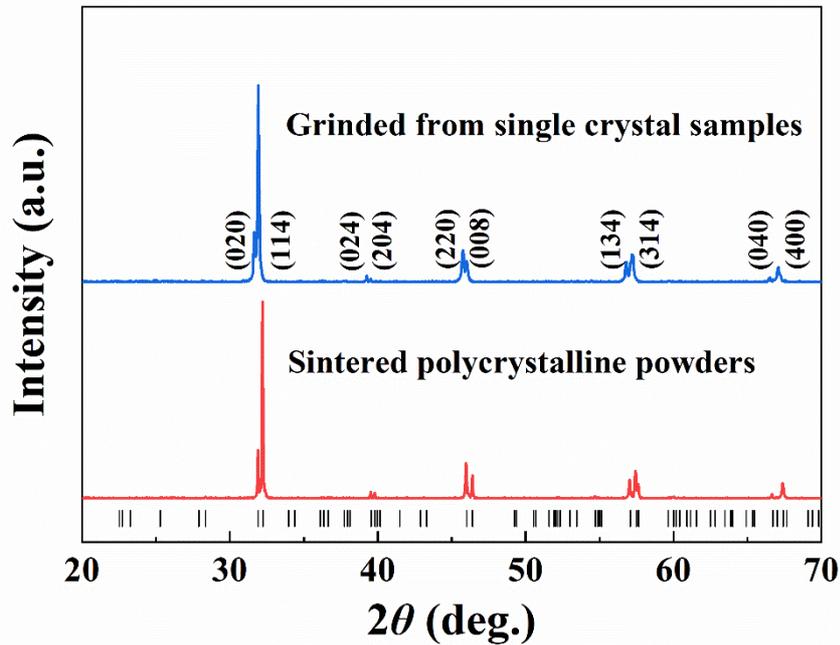

**Figure. 1. X-ray diffraction patterns of grinded ANO single crystal sample and sintered polycrystalline sample.**

TEM experiment was also performed to reveal the sample quality and the ferroic domain structure. Fig. 2(a) and (b) present the representative domain patterns and domain walls in ANO single crystal. The selected area electron diffraction (SAED) patterns of two adjacent domains are acquired and shown in Fig. 2(c) and (d), respectively. Clearly, the zone axis has a 90° rotation from one domain to the other domain, i.e. 90° ferroelastic domain wall. Moreover, the (*00l*) reflections with $l = 2n + 1$ (003 are marked by red arrow) in Fig. 2(d) give the evidence of the existence of polar phase with space group *Pmc*2$_1$.[1,13] Besides, the dark-field images using the (003) reflection reveal that there exists high density of antiphase-boundaries in ANO single crystal, as shown in corresponding inset images for the enlarged regions in Fig. 2 e and f. These antiphase-boundaries are caused by the antiparallel cation displacement, which are also observed in PbZrO$_3$ single crystal and ANO based ceramics.[40,41]

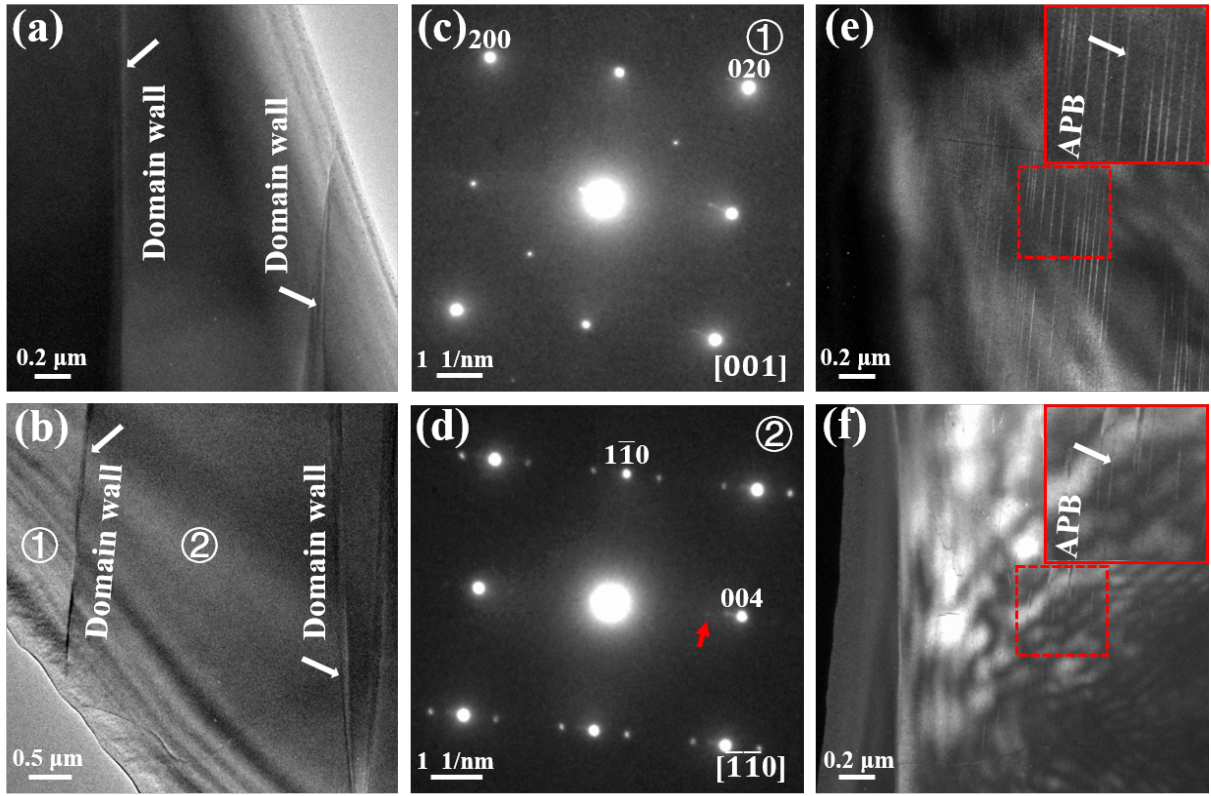

**Figure 2. TEM characterization of ANO single crystal.** (a-b) Overview TEM images taken from different areas from ANO single crystal, the white arrow points to where domain walls are; (c-d) SAED patterns taken from domains labeled ①, ② in (b), respectively. The (003) reflection is marked by red arrow. (e-f) Dark-field images taken from different areas from ANO single crystal, the red box is an enlarged view of the dotted box and the white arrow points to where antiphase-boundaries (APBs) are.

Domain morphology and domain walls in ANO single crystal can be clearly seen in Fig. 3(a). Moreover, to exclude double diffraction, we gradually tilt the sample while keeping the (*00l*) direction (Fig. 3(b-e)). When the sample was tilted near two-beam condition (Fig. 3(e)), the paths for double diffraction should not exist. In this case, we can see the (003) reflections as well, which suggests that the (003) reflections are caused by the structure itself. Therefore, the above results confirm that the space group of ANO single crystal is *Pmc*2$_1$.

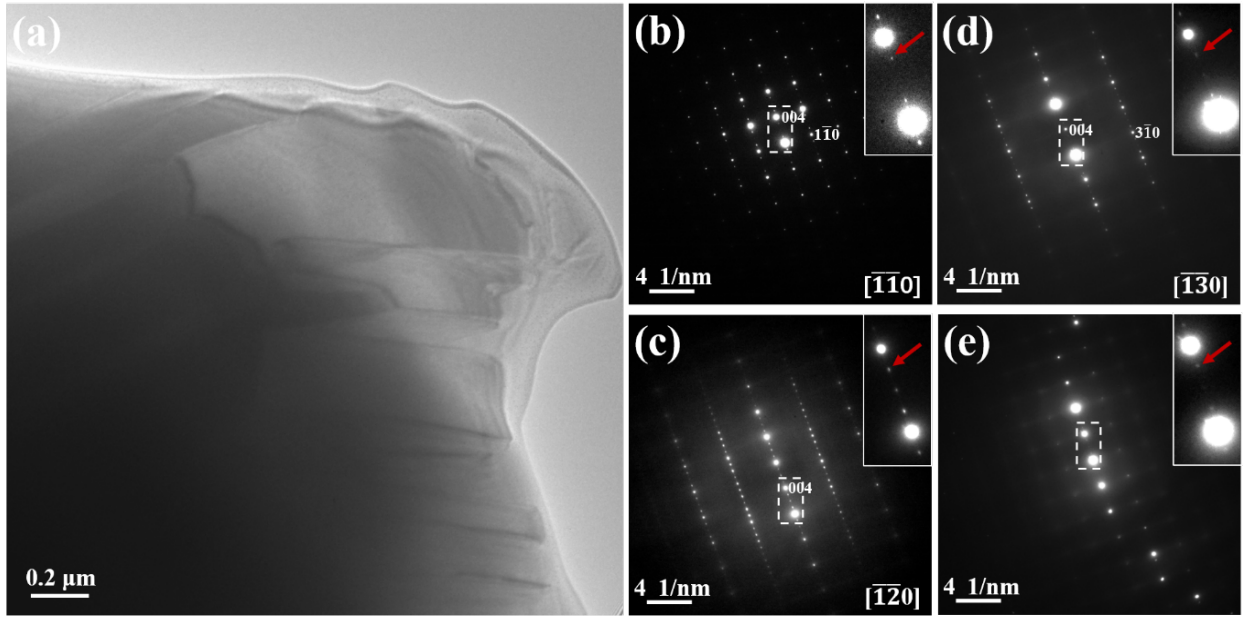

**Figure 3. TEM characterization of ANO single crystal.** (a) Overview bright-field image of ANO single crystal; (b-e) A series of SAED patterns obtained from ANO single crystal, the white dash boxes are magnified to clearly see the 003 reflections (red arrows).

To investigate the ferroelastic domain structure of ANO single crystal,[16,42,43] we also performed the reflection-mode angular-resolved polarized light microscopy measurement and observed the ferroelastic domains, as shown in Fig. 4. Fig. 4(a), (e) and (i) are obtained in reflection-mode with the sample rotation angles of 0°, 45° and 90°. The identical blue-boxed region was thoroughly investigated by using polarized light. Stripe-like domain patterns with bright and dark contrasts indicate different ferroelastic domains, the yellow arrows represent the ferroelastic directions. As the analyzer rotated at a suitable angle, the domain contrast completely reversed, signifying the ferroelastic nature, as shown in Fig. 4(b), (c) and (d). The sample was then rotated 45° and 90°. It can be clearly seen that the identical contrast of neighboring domains in Fig. 4(f) and (h), but there are inverse contrasts in Fig. 4(b), (d) and Fig. 4(j), (l). This phenomenon can be explained by the similar and different intensity of the reflected light passing through the analyzer and we can define the ferroelastic domain wall in ANO single crystal to be 90°. Meanwhile, sample rotation relative to the polarizer-analyzer pair can identify an extinction angle($\theta_e$, at which bright domains turn dark),[44] as shown in Figs. 4(c), (g) and (k). It can be seen that the polarizer-analyzer pair remains perpendicular to each other, but ferroelastic domains are indistinguishable and domain walls are blurry and barely visible in Fig. 4(c) and (k), while clearly contrast of adjacent domains was observed in Fig. 4(g). Based on Fig. 4(c), (g) and (k), the extinction angle θ_e=90°, as expected for the orthorhombic phase. Simultaneously, the observed surfaces in PLM are more likely in {100} planes, due to the perpendicular edges the crystal has, which indicates the mutually orthogonal crystal lattice directions in orthorhombic structure.

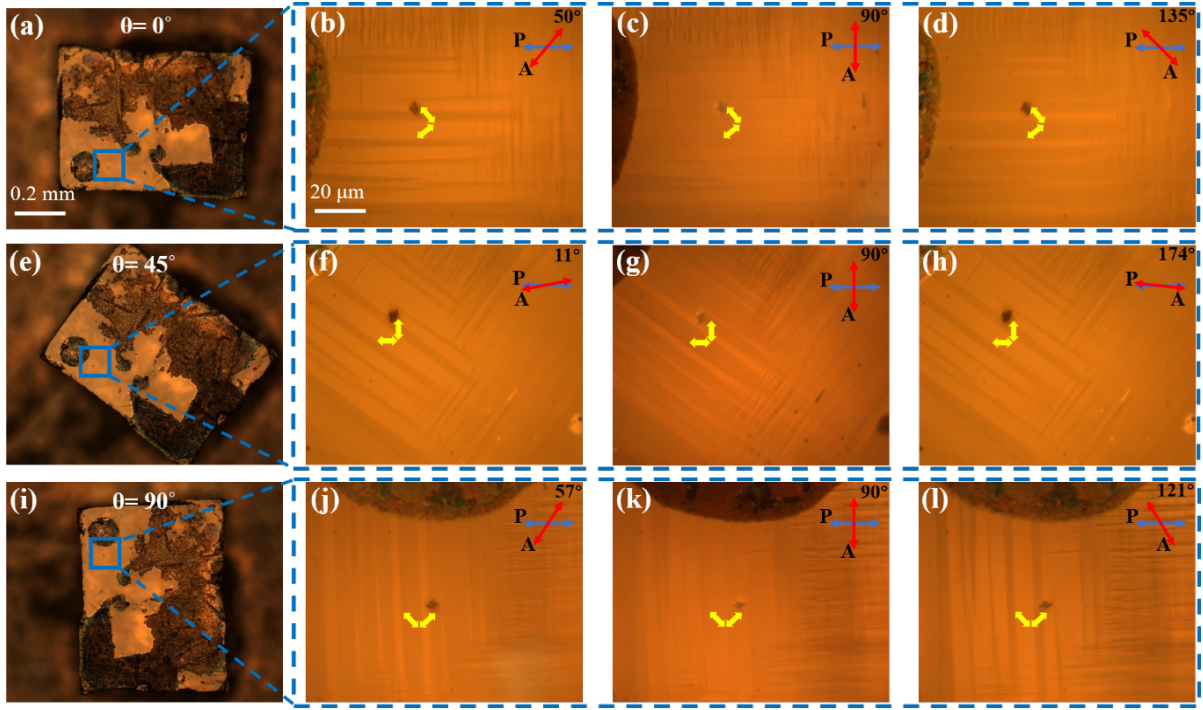

**Figure. 4. Ferroelastic domain structures of reflection-mode angular-dependent polarized light microscope on ANO single crystal.** (a) show the optical image of the as grown ANO single crystal with sample rotation angle θ=0°; (b-d) show the enlarged optical image in the blue boxed region with different polarizer-analyzer angles. (e-h) and (i-l) show the corresponding reflective optical images with the sample rotation angle θ=45° and 90°, respectively. The red and blue arrows indicate the directions of analyzer and polarizer, respectively. The yellow arrow represents the directions of ferroelastic domains in ANO single crystal.

We then performed PFM in the identical region as shown in polarized optical image to further investigate the domain structures observed under polarized optical microscopy. Fig. 5(a) is an optical image in reflection mode, in which the domain structures are absent, only topography is observed. By switching to the polarized light mode, the contrast of neighboring domains can be clearly seen and be reversed by rotating the analyzer. PFM data are displayed in Fig. 5 (d-f), revealing the same domain structure as the optical images show. The surface shows deep rill-like folds, and has no crosstalk with the corresponding PFM signals. We clearly observed an out-of-plane electromechanical signal, with an obvious phase contrast indicates the existence of net polarization in ANO single crystal caused by the antiparallel cation displacement along the *c* axis[23] from polar ferrielectric space group *Pmc*2$_1$. The amplitude image also exhibits strong signal contrast for two adjacent domains, which is also found in other ferroic systems with coexistence of ferroelastic and ferro/ferrielectric domains.[44–46] The domain morphology observed by PLM and PFM are in good agreement with what have been observed by using dark-field TEM. It can be clearly seen that the stripe-like domain patterns by TEM has similar characteristic as the one in PLM and PFM. Combining the TEM, PLM and PFM measurements, we can come to a conclusion that due to the polar space group *Pmc*2$_1$, the ANO single crystal has a net polarization, and the corresponding domain morphology detected by the PFM has similar pattern with the domain observed by TEM and PLM, suggesting the coexistence of ferroelastic domain and ferrielectric domains.

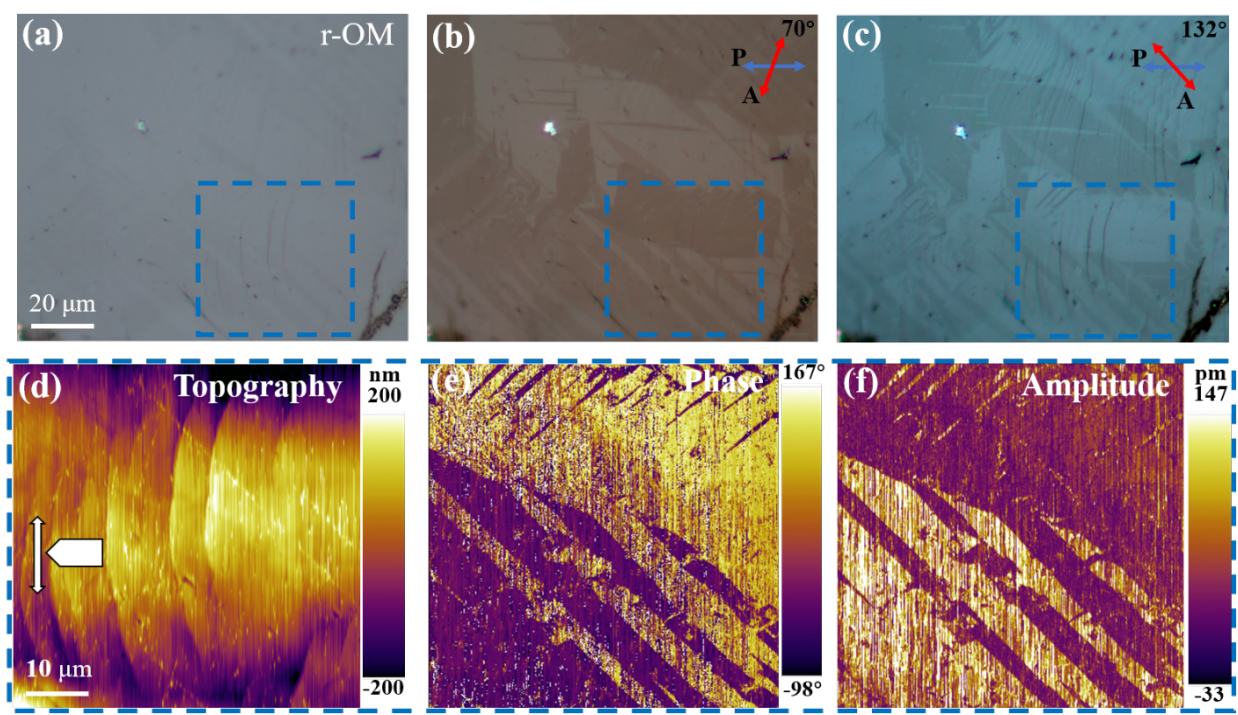

**Figure 5. Domain structures of ANO single crystal by PLM and PFM.** (a) Reflective optical microscope images of sample surface without polarized light; (b) and (c) Polarized light microscope images at identical region as shown in (a); (d-f) PFM characterization of ANO single crystal taken from blue dashed box region: the white arrow indicates the scan direction; (d) is AFM topography, which is consistent with the optical image (a); (e) and (f) show the out-of-plan

phase and amplitude images.

In summary, we synthesized ANO single crystal by the conventional flux method. A pure phase of ANO structure is verified by XRD. By utilizing TEM measurement, we clarify that the space group is *Pmc*2$_1$, which cannot be distinguished from XRD data. In addition, TEM data also suggest a ferrielectric phase, and the polar nature gives the ferroelastic and ferrielectric domains, which coexist in ANO. The coexistence is further experimentally verified by PLM and PFM measurement. The observation of ferroic domains in ANO suggests that ANO single crystal is in polar phase. Our results offer an opportunity to study the ferroic domains in other ANO based systems, such as Tantalum or Samarium doped single crystals, and may suggest domain engineering strategy to enhance the energy storage performance.

**Acknowledgements**

The work was supported by National Natural Science Foundation of China (Grant Nos. 11572040, 11604011 and 51972028), the National Key Research and Development Program of China (No. 2019YFA0307900), and Beijing Natural Science Foundation (Grant No. Z190011). X.W. also acknowledge the technological Innovation Project of Beijing Institute of technology.